\documentclass[aps,a4paper,preprint]{revtex4-1}
\pdfoutput=1
\usepackage{graphicx}   % need for figures
\usepackage[skip=1pt,font={sf,footnotesize}]{caption}
\usepackage{color}      % use if color is used in text
\usepackage{subcaption} %use for side-by-side figures; for subfigure use, DONOT use subfig or subfigure package, built in subfigure usage

\usepackage{hyperref}   % use for hypertext links, including those to external documents and URLs
\usepackage{amsmath}
\numberwithin{equation}{section}
\usepackage{amssymb}
\usepackage{verbatim}
\usepackage{float}
\usepackage{natbib}
\usepackage{titlesec}

\makeatother

\usepackage{array} 
\renewcommand{\arraystretch}{1.5}   
\usepackage{tabularx}
\usepackage{booktabs}
\usepackage[utf8]{inputenc}
\usepackage{bm}

\begin{document}
\title{Modeling sRNA-regulated Plasmid Maintenance}
\author{Chen Chris Gong}
\email[]{cgong@uni-potsdam.de}
%\homepage[]{Your web page}
%\thanks{}
\affiliation{Max Planck Institute of Colloids and Interfaces, Science Park Golm, 14424 Potsdam, Germany}
\affiliation{Institute of Physics and Astronomy, University of Potsdam, Karl-Liebknecht-Stra\ss e 32, 14476 Potsdam, Germany}

\author{Stefan Klumpp}
\email[]{stefan.klumpp@phys.uni-goettingen.de}
\affiliation{Max Planck Institute of Colloids and Interfaces, Science Park Golm, 14424 Potsdam, Germany}
\affiliation{Institute for Nonlinear Dynamics, Georg August University of G\"ottingen, Friedrich-Hund-Platz 1, 37077 G\"ottingen, Germany}

\date{\today}

\begin{abstract}
We study a theoretical model for the toxin-antitoxin (hok/sok) mechanism for plasmid maintenance in bacteria. Toxin-antitoxin systems enforce the maintenance of a plasmid through post-segregational killing of cells that have lost the plasmid. Key to their function is the tight regulation of expression of a protein toxin by an sRNA antitoxin. Here, we focus on the nonlinear nature of the regulatory circuit dynamics of the toxin-antitoxin mechanism. The mechanism relies on a transient increase in protein concentration rather than on the steady state of the genetic circuit. Through a systematic analysis of the parameter dependence of this transient increase, we confirm some known design features of this system and identify new ones: for an efficient toxin-antitoxin mechanism, the synthesis rate of the toxin's mRNA template should be lower that of the sRNA antitoxin, the mRNA template should be more stable than the sRNA antitoxin, and the mRNA-sRNA complex should be more stable than the sRNA antitoxin. Moreover, a short half-life of the protein toxin is also beneficial to the function of the toxin-antitoxin system. In addition, we study a therapeutic scenario in which a competitor mRNA is introduced to sequester the sRNA antitoxin, causing the toxic protein to be expressed. 
\end{abstract}

\maketitle

\setcounter{secnumdepth}{3}

\renewcommand\thesubsection{\arabic{subsection}}

\section{Introduction} \label{intro}

Small regulatory RNA (sRNA) plays an important role in gene regulation in organisms from bacteria to mammals by controlling for example translation and/or mRNA stability and the list of known RNA regulation systems keeps increasing at a rapid pace \cite{Altuvia, Waldor,Waters2009615}. sRNA regulation possesses characteristics that are distinct from protein regulation, in particular a threshold-linear response, which provides an ultrasensitive mechanism for regulatory switching, and the possibility of hierarchical crosstalk, which allows prioritizing of expression \cite{levine07}. 

Some of the best known sRNA regulation systems are related to plasmid replication and plasmid maintenance in bacteria. In the first case, an sRNA controls whether synthesis of a replication primer proceeds to plasmid replication \cite{Summers, Paulsson01, Eguchi91}. In the second, the plasmid encodes a (type I) toxin-antitoxin system such as the hok/sok system (``host-killing/supression-of-killing''), encoding a protein toxin and an antisense RNA which acts as an antitoxin by binding to the toxin mRNA and blocking ribosome access, thus preventing toxin synthesis \cite{Wagner07}. (Other types of toxin-antitoxin systems have different functions \cite{Magnuson}, in particular related to the formation of persister cells \cite{Lewis}.)

\begin{figure}[ht]
 \includegraphics[scale=0.55]{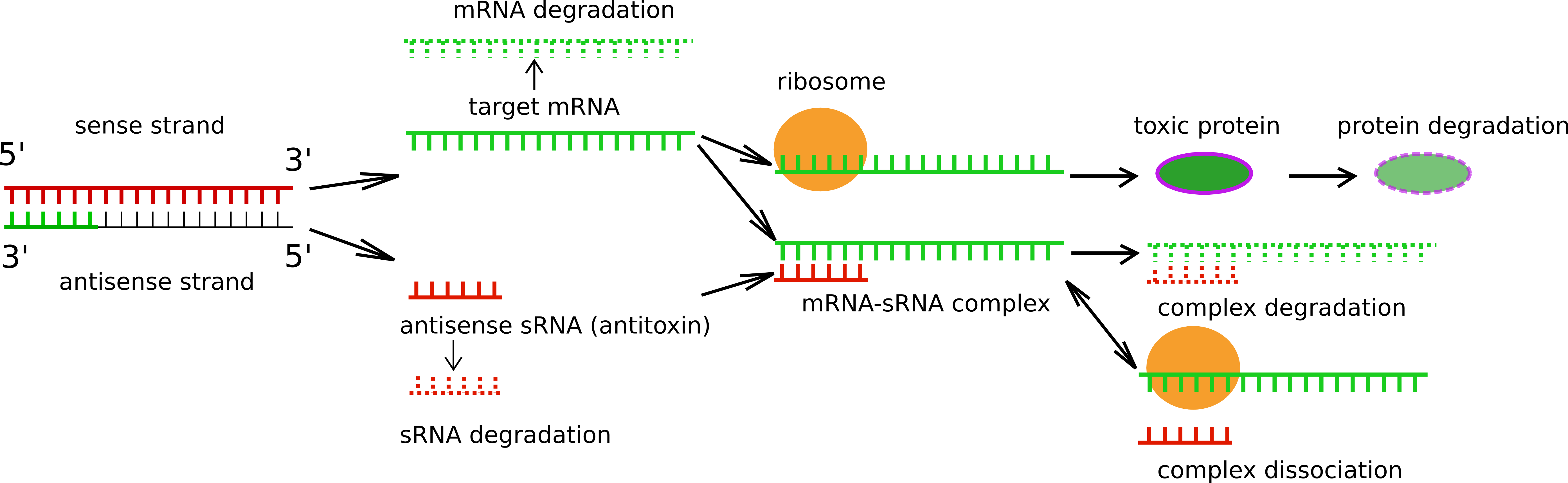} 
 \caption{Simplified drawing of the sRNA regulated toxin-antitoxin mechanism for plasmid maintenance. Structural transitions and processing of the mRNA have been omitted \cite{Wagner07}. \label{intro1}}
\end{figure}

The toxin-antitoxin system, which is also known as an ``addiction module'', maintains the plasmid number through post-segregational killing of plasmid-free progeny due to differential stability of the toxin and antitoxin RNAs. The killing is done by a potent protein toxin that irreversibly damages the cell membrane \cite{Gerdes86}. In a steady state with a stable plasmid concentration, sRNA antitoxin exists in excessive molar amount compared to target mRNA, such that the latter is entirely sequestered in translationally inactive sRNA-mRNA complexes \cite{Gerdes92}. However, when a progeny cell becomes plasmid-free through cell division, synthesis of both toxin mRNA and antitoxin sRNA are stopped. The sRNA, which has a very short half-life, is rapidly depleted and the more stable target mRNA can be translated into toxic protein, killing the cell (Figure \ref{intro1}). 

To understand the design of the gene circuit encoding the toxin-antitoxin mechanism of post-segregational killing, here we analyze a theoretical model for the dynamics of such a circuit. The model is related to and extends previous models for sRNA-based post-transcriptional regulation \cite{levine07,Legewie08,Mitarai09}. It allows us to address the parameter dependence of the genetic circuit to identify essential features and criteria for its efficient function, for example, whether there are any criteria beyond the difference between toxin and antitoxin RNA lifetime. It also allows us to test whether the system can be designed in such a way that only complete loss of the plasmids triggers killing and not a low but non-zero plasmid copy number. We also use the model to study a proposed antibacterial strategy \cite{Williams,Faridani} making use of regulatory crosstalk between RNAs. To answer these questions we performed extensive parameter sweeps, scanning all parameters of the model individually as well as extensively testing randomly chosen parameter sets, thus varying all parameters simultaneously. Overall, the model confirms the known principles for the function of the toxin-antitoxin mechanism as the dominant ones, in particular, the difference in RNA lifetime and the threshold condition on the synthesis rates. In addition, it also indicates a few new design features. Specifically, our analysis of the parameter dependence shows that the stability of the mRNA-sRNA complex plays an important role as well and should exceed the stability of the free antitoxin. Moreover, an unstable toxic protein is superior to a stable one for induction of killing upon plasmid loss.

\clearpage
\section{Model and Methods}
 
\subsection{Dynamic Equations and Analytical Solution of Steady State Concentrations} \label{sec: secss}

The dynamics of a toxin-antitoxin system is modelled with four coupled ordinary differential equations for four dynamical variables, the concentrations of the toxin mRNA ($m$), the antitoxin sRNA ($s$), the toxic protein ($p$) and of the mRNA-sRNA complex ($c$), in which the mRNA is silenced. All concentrations are expressed in units of number of molecules per volume of a cell. The four equations describe the synthesis and degradation of the mRNA, the sRNA, and the protein, as well as the formation and dissociation of the mRNA-sRNA complex:
\begin{subequations} \label{eq:1}
\begin{align}
 \dot m & = \alpha_{m} \cdot g - \beta_{m} \cdot m - h^{+} \cdot m \cdot s + h^{-} \cdot c \\
 \dot s & = \alpha_{s} \cdot g - \beta_{s} \cdot s - h^{+} \cdot m \cdot s + h^{-} \cdot c \\
 \dot c & = h^{+} \cdot m \cdot s - h^{-} \cdot c - \beta_{c} \cdot c \\
 \dot p & = \alpha_{p} \cdot m - \beta_{p} \cdot p
 \end{align}
\end{subequations}

The $10$ parameters of these equations are as follows: $\alpha_{m}$, $\alpha_{s}$, $\alpha_{p}$ are the synthesis rates of mRNA, sRNA and protein (transcription and translation rates, respectively); $\beta_{m}$, $\beta_{s}$, $\beta_{c}$, $\beta_{p}$ are the degradation rates of mRNA, sRNA, the mRNA-sRNA complex and protein, respectively; $h^{+}$ and $h^{-}$ are the binding and unbinding rate of the complex, and $g$ is the plasmid copy number per cell volume.

The steady state solution is obtained by setting the time derivatives on the left hand side of the equations to zero. The steady state for this system can be explicitly given, because the nonlinear terms cancel each other in equations (1a) and (1b), leaving a quadratic equation which can be solved analytically. This leads to the steady state sRNA concentration:
\begin{equation}
\vspace{-5mm} 
s^{*} = \dfrac{ -A \pm \sqrt{A^{2} + 4 \beta_{s}\beta_{m}\alpha_{s} g \dfrac{ h^{+} \beta_{c} }{ h^{-} + \beta_{c} } } } {2 \beta_{s} \dfrac{ h^{+} \beta_{c} }{ h^{-} + \beta_{c} } } \hspace{3mm}, \nonumber \\ 
\end{equation}
where
\begin{equation}
A = (\alpha_{m} g - \alpha_{s} g) \dfrac{h^{+}\beta_{c}}{h^{-} + \beta_{c}} + \beta_{s}\beta_{m}\hspace{3mm}. \nonumber \\ 
\end{equation}
The steady state concentrations of the other three components can be in turn given as:
\begin{equation}
\begin{aligned}
m^{*} & = \dfrac{\alpha_{m}g - \alpha_{s}g + \beta_{s}s^{*}}{\beta_{m}} \\
c^{*} & = \dfrac{m^{*} s^{*} h^{+}}{h^{-}+\beta_{c}} \\
p^{*} & = \dfrac{\alpha_{p} m^{*}}{\beta_{p}} \label{eq:pstar} \\
\end{aligned}
\end{equation}

This result includes some limiting cases that have been studied in earlier work on sRNA-dependent gene regulation. In the limit of large binding and unbinding rates $h^{+}$ and $h^{-}$ (i.e. the limit of ``rapid equilibrium''), the steady state concentration of the RNA complex concentration becomes $c^{*} = (m^{*}s^{*}h^{+})/h^{-}$, as previously obtained by Legewie et al. \cite{Legewie08}. The limit of irreversible binding, $h^{-} = 0$, in which only the equations for sRNA and mRNA concentrations need to be considered, was previously studied by Levine et al. \cite{levine07} and Mitarai et al. \cite{Mitarai09}. 

\subsection{Numerical Methods}

To study the dynamic behavior, equations \ref{eq:1}(a-d) are numerically integrated using the 4th order Runge-Kutta method with an integration time step on the order of $1\times10^{-4}$ min, for a time span of 300 min. We note that the time unit of the dynamics could be made dimensionless by rescaling all rates relative to one rate that determines the time unit. Correspondingly, the results presented below will typically depend on ratios of time scales or rates. To achieve both stability and speed, an adaptive time step for the numerical integration is implemented, such that when the integration becomes numerically unstable, a smaller time step is used. A tell-tale sign for numerical instability is the appearance of negative concentration values of one or more components. The analytical results for the steady state concentrations are in good agreement (up to floating point error) with the results obtained from the numerical integration. 
\section{Results and Discussion}

\subsection{Qualitative Description of the Dynamics of the Genetic Circuit} \label{sec: num}

\begin{figure}[hb!]
 \includegraphics[scale=0.5]{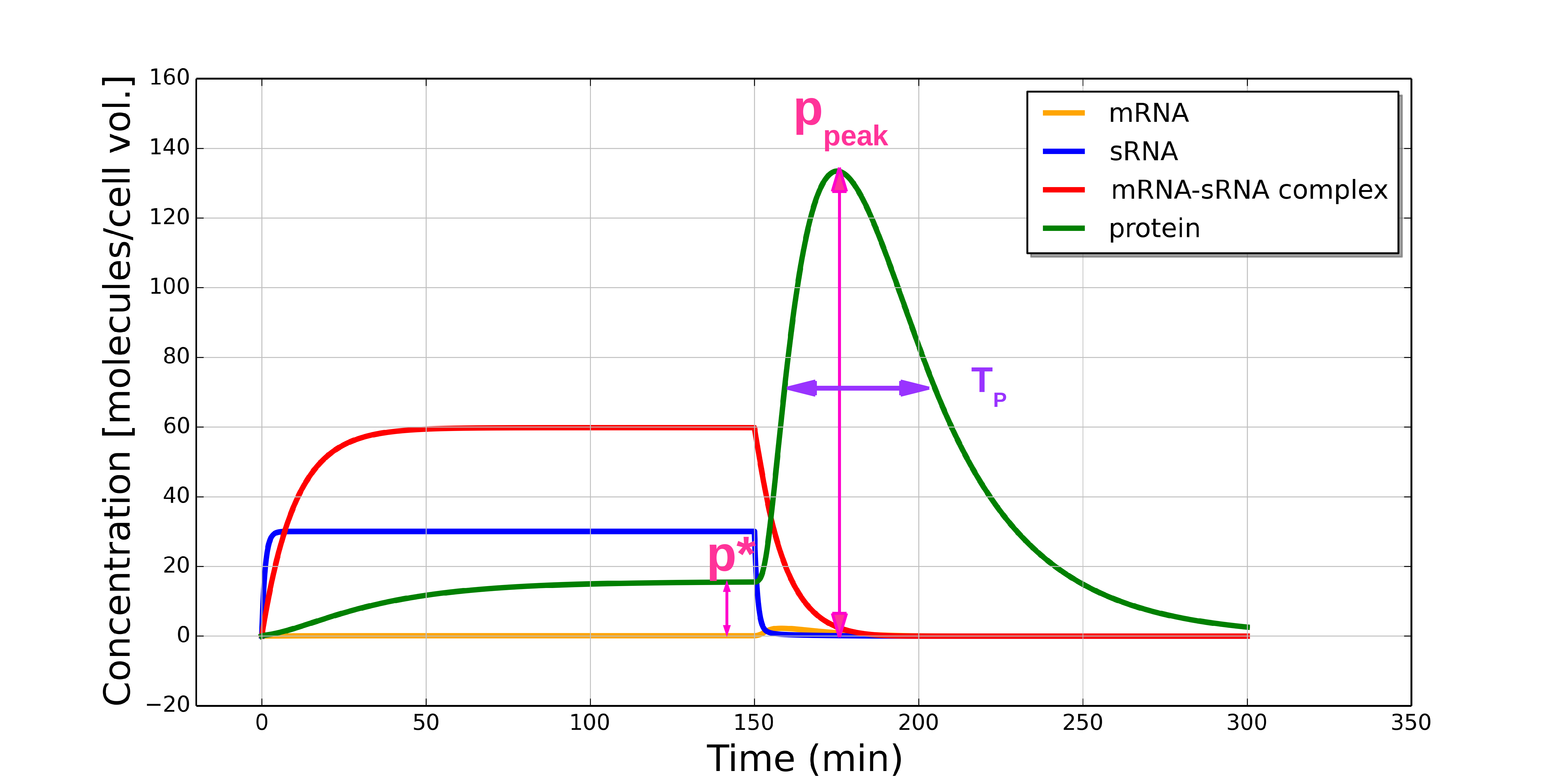} 
 \caption{Simulated dynamics of the toxin-antitoxin gene circuit: The different colors show the concentrations (in molecules per cell volume) of mRNA (orange), sRNA (blue), mRNA-sRNA complex (red) and protein (green). At $t = 150$ min, the plasmid is lost, described by resetting the copy number $g$ to zero, inducing a transient peak in the protein concentration. The protein concentration peak fold-increase $R$ is $8.6$ and the peak width $T_{p}$ equals $48$ min. The parameters are as listed in the last row of table \ref{tab:Tablelit} and $g = 6.0$, $h^{+} = 20.0$, $h^{-} = 1.0$, $\beta_{c} = 0.1$. \label{fig10real} }
\end{figure}

For the model described above, the following scenario is considered: the dynamics of the toxin-antitoxin system begin with a cell which has just acquired one or more copies of the plasmid, but has not yet synthesized any of its products, i.e. we start with $m=s=c=p=0$. The dynamics are numerically integrated for a sufficiently long time, so that a steady state of the system is reached. At a certain time point (here at $t=150$ min of a total $t = 300$ min), the plasmid copy number is set to $0$ to mimic plasmid loss, e.g. due to insufficient replication or unequal partitioning of the plasmid during cell division. To illustrate the dynamics, realistic values of the parameters are used, which are estimated from the literature (Table \ref{tab:Tablelit}). Equations (\ref{eq:1}) are numerically integrated to produce Figure \ref{fig10real} .

\begin{table}[ht!]
\footnotesize
\renewcommand{\arraystretch}{0.8}
\begin{tabular}{|p{3cm}|p{2.2cm}|p{2.6cm}|p{2.2cm}|p{2cm}|p{2.2cm}|p{1.8cm}|}
\hline
 & $\alpha_{m}$[gene copy$^{-1}$min$^{-1}$]& $\beta_{m}$[min$^{-1}$] & $\alpha_{s}$[gene copy$^{-1}$min$^{-1}$] & $\beta_{s}$[min$^{-1}$] & $\alpha_{p}$[min$^{-1}$] & $\beta_{p}$[min$^{-1}$]\\
 \hline
 Range of values & 0.1-15.8 & 0.034-0.693 & 0.1-15.8 & 0.35-1.4 & 1-7 & 0.0346 \\ \hline
References & \cite{levine07,Liang99,Linchao87} &  \cite{gg90,berstein02} & \cite{levine07,Liang99,Linchao87,wagner02} & \cite{Jas,Linchao91,Wagner07,gg90} &  &  \cite{Wagner07} \\
\hline
Default parameters used here  & 1.0 & 0.2 & 6.0 & 1.0 & 5.0 & 0.035 \\
\hline
\multicolumn{7}{l}%nextline leave blanck

\end{tabular}
\caption{Table of parameter values based on the literature for toxin-antitoxin systems and related sRNA regulation systems. For some parameters, the values are know to depend on growth conditions. For components with long half life, the degradation rate is dominated by dilution through cell growth and division.\label{tab:Tablelit}}
\end{table}

In Figure \ref{fig10real}, all four concentrations increase initially until the first steady state is reached. At $t = 150$ min, due to the loss of all plasmid copies, the synthesis of new RNA molecules stops. Nevertheless, a transient increase in the free mRNA concentration is observed, which results in a transient increase of the protein concentration. The transient increase in mRNA is a result of sRNA degrading much faster than mRNA, so when a mRNA-sRNA complex dissociates, there is a surplus amount of free mRNA molecules released, which in turn allows the synthesis of toxic protein. Thus, the main function of the toxin-antitoxin system, plasmid maintenance via the synthesis of a toxin upon plasmid loss, which results in the removal of plasmid-free cells from the population, is dependent on a transient dynamics rather than on a steady state. This behavior is in contrast to other sRNA-based regulation systems, which are based on similar mechanisms, but control the steady state concentration of mRNA \cite{levine07}.

For a quantitative characterization of the dynamics, specifically that of the protein concentration change after the loss of plasmids, two quantities are measured for each simulated scenario: (i) 
the peak fold-increase of the protein concentration, $R$, defined as the ratio of the maximum of the protein concentration after plasmid loss and the steady state protein concentration before the plasmid loss, $R=p_{\rm peak}/p^*$; and (ii) the transient peak width $T_{p}$, defined as the width of the peak at half maximum. 

The effectiveness of the toxin-antitoxin mechanism is partially determined by the protein concentration peak fold-increase $R$, as opposed to by the absolute concentration of the toxin. A high peak fold-increase allows a toxicity threshold to be set, such that the fluctuations of the steady state protein concentration are very unlikely to cross such a threshold and thereby ``accidentally'' lead to cell death, while simultaneously permitting the transient increase of protein concentration when the plasmid number is not being properly maintained to easily overcome such a threshold. For example, if the cell will be killed by a $10$-fold-increase in toxin concentration ($10$ being the killing threshold), and if there is little chance of an accidental $10$-fold protein concentration increase due to other factors and noise, the dynamics which can produce $R > 10$ are considered effective in exerting a robust control on the host cell. 

\subsection{Dependence of the protein dynamics on individual model parameters} \label{sec: spv}

We first investigate the parameter dependence of the model by varying each parameter in isolation, holding the other 9 fixed at constant values. In all cases, the peak fold-increase $R$ and the peak width $T_{p}$ are determined as functions of the modulated parameters. 

\begin{figure}[hb]
\captionsetup[subfigure]{justification=centering}
\centering
\includegraphics[width=1\textwidth]{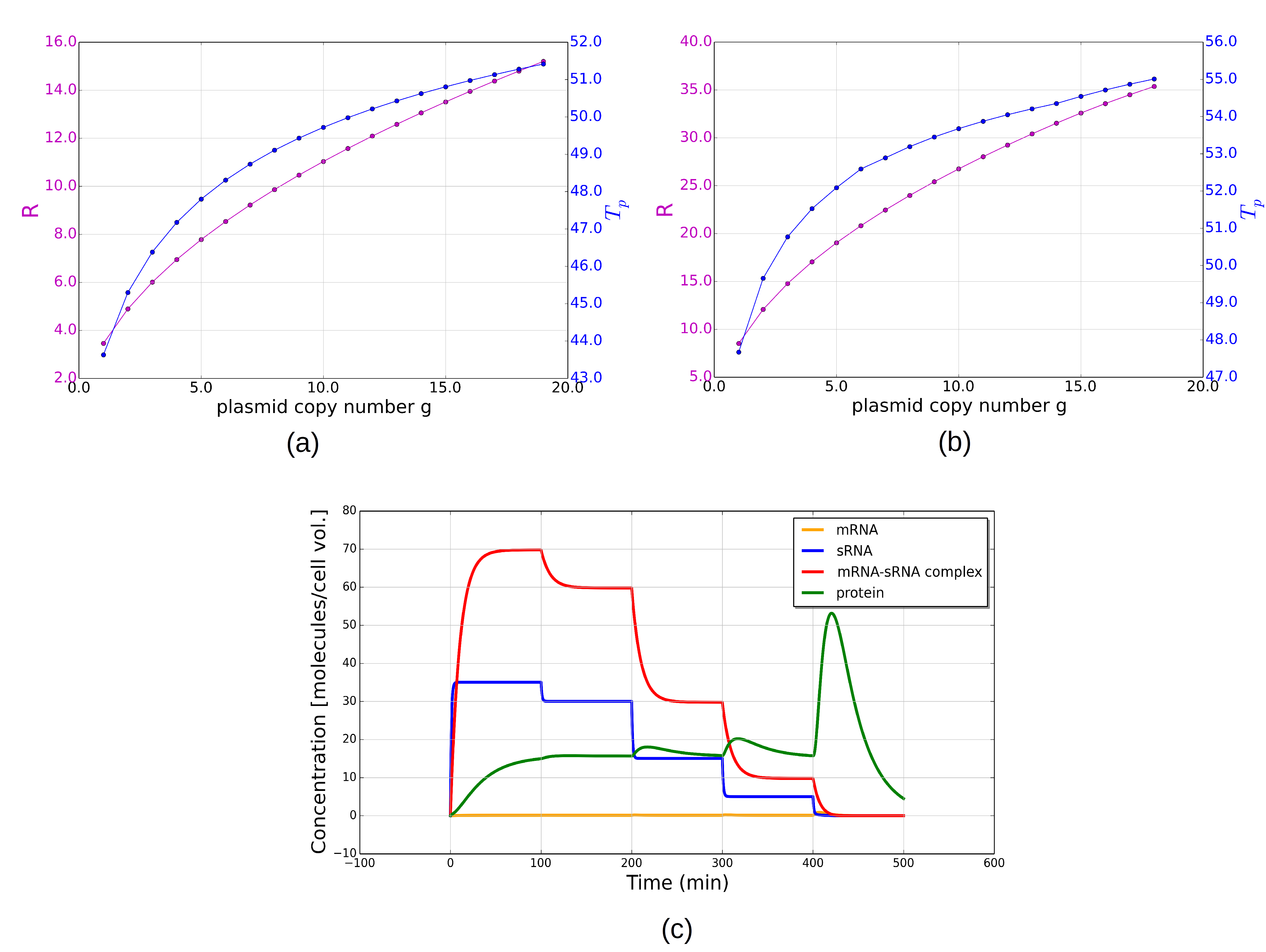}
\caption{Dependence on plasmid copy number $g$: (a,b) protein concentration peak fold-increase $R$ and width $T_{p}$ as a function of the plasmid copy number $g$ for our default binding and unbinding rates $h^{+} = 20, h^{-} = 1$ (a), and for rapid equilibrium with $h^{+} = 1000, h^{-} = 10$ (b). (c) the simulation of sequential plasmid loss under default parameters. The plasmid copy number is reduced from $7 \rightarrow 6 \rightarrow 3 \rightarrow 1 \rightarrow 0$. A substantial increase in toxin concentration only occurs when the last plasmid is lost. \label{figgscan}}
\end{figure}

First, the dependence of $R$ and $T_{p}$ on the plasmid copy number $g$ (Figure \ref{figgscan}) is investigated. An increase of $g$ is equivalent to increasing the synthesis rates of the sRNA and mRNA molecules by the same ratio. $R$ and $T_{p}$ are both positively affected when $g$ is increased. For the default parameters (Figure \ref{figgscan} (a)) as well as for the rapid equilibrium case (Figure \ref{figgscan} (b)), the rates of change of both $R$ and $T_{p}$ (i.e. the slope of the curves) are higher when $g$ is low, and lower when $g$ is high. In the case of rapid equilibrium, i.e. when both $h^{+}$ and $h^{-}$ are high, the changes in $R$ and $T_{p}$ at the loss of one or two plasmid copies are more substantial than in the cases of lower binding and unbinding rates. This is to say, in the case of rapid equilibrium, the model could be effective in triggering killing even when the plasmids are not completely lost.

Another way to think about plasmid maintenance is shown in the simulation of sequential loss of plasmid copies under default parameters (Figure \ref{figgscan} (c)). The change in plasmid number does not result in the sudden release of large amount of toxin until the last plasmid is lost. Comparing to when all plamids are lost at once (Figure \ref{fig10real}), sequential plasmid loss will result in a less prominent transient peak when the last plasmid is lost. This observation is consistent with the function of the toxin-antitoxin system to enforce plasmid maintenance, i.e. the host cells are forced to retain at least one copy of the plasmid, but there is little dependence on whether there are more or fewer copies. 

\begin{figure}[hb]
\captionsetup[subfigure]{justification=centering}
\centering
\includegraphics[width=1\textwidth]{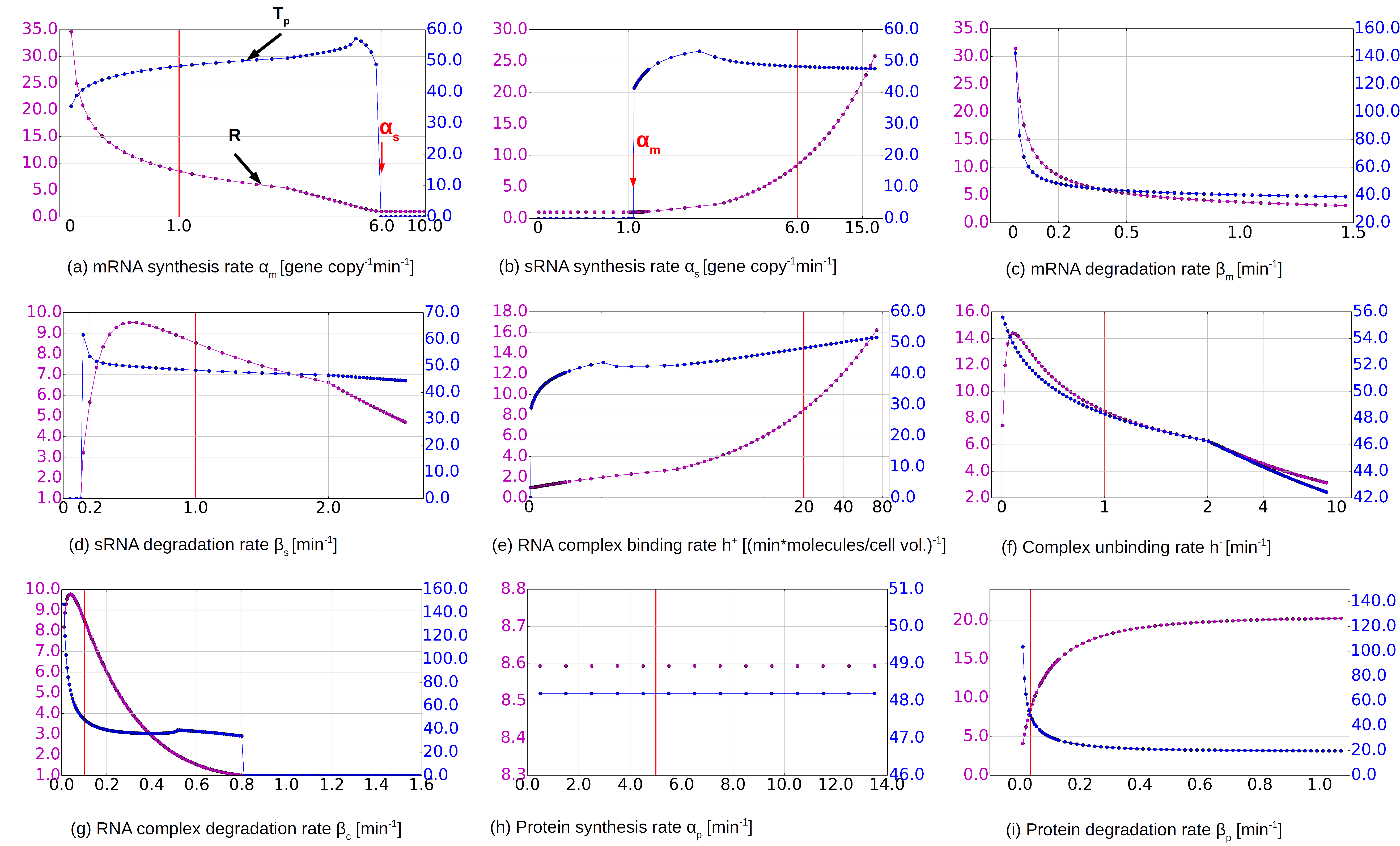}
\caption{\label{fig:singlescan} Dependence on individual model parameters: Each model parameter is varied in isolation while the other 9 parameters are held fixed. The toxic protein concentration peak fold-increase $R$ (purple) and the width of the peak $T_{p}$ (blue) are plotted as functions of each parameter in the 9 subfigures. The default parameter values (from Table \ref{tab:Tablelit} and as in Figure \ref{fig10real}) are marked by red lines in the plots.}
\end{figure}

The dependence of $R$ and $T_{p}$ on the other 9 parameters are shown in Figure \ref{fig:singlescan} .
In general, two conditions for the existence of a transient peak of the toxic protein concentration can be extracted from these parameter dependence: 

First, the synthesis rate of sRNA ($\alpha_{s}$) must exceed the synthesis rate of mRNA ($\alpha_{m}$). This threshold condition can be observed by comparing the varying values to the default values for mRNA and sRNA synthesis rates. In Figure \ref{fig:singlescan}(a), when the synthesis rate of mRNA $\alpha_{m}$ exceeds the default synthesis rate of sRNA $\alpha_{s}$ at 6/min/gene copy, the peak disappears. Conversely in Figure \ref{fig:singlescan}(b), when $\alpha_{s}$ becomes larger than $\alpha_{m}$ at a default value of 1/min/gene copy, the protein concentration peak starts to appear ($R > 1$).

This threshold condition is well-known for sRNA regulation \cite{levine07}. It can be explained as follows. Increasing the number of mRNA molecules while keeping the number of sRNA molecules constant increases both the protein concentration peak $p_{peak}$ as well as the first steady state protein concentration $p^{*}$. However, $p^{*}$ increases faster than $p_{peak}$ as $\alpha_{m}$ increases. Therefore the relative increase of the protein concentration $R$ diminishes as $\alpha_{m}$ increases. A high level of free mRNA before plasmid loss will also mean that plasmid-containing cell might ``poison itself'' without any loss of plasmids. When $\alpha_{m}>\alpha_{s}$, there will be more free mRNA than what can be ``neutralized'' by complex formation, and the toxic protein will be expressed at a high level before the loss of plasmids. The toxin-antitoxin mechanism relies on the low amount of surplus of free mRNA after the loss of plasmids, so that when $\alpha_{m}$ is higher than $\alpha_{s}$, this effect is lost.

Second, the degradation rate of sRNA must exceed the degradation rate of mRNA: $\beta_{s} > \beta_{m}$. This condition ensures that after the loss of plasmids, a pool of free mRNA builds up, since the sRNA is degraded more rapidly. When sRNA is more long-lived than mRNA, that is, when $\beta_{m}$ exceeds $\beta_{s}$ at 1/min, or when $\beta_{s}$ is lower than $\beta_{m}$ at 0.2/min (Figures \ref{fig:singlescan} (c), \ref{fig:singlescan} (d)), a pronounced protein concentration peak is not observed. We notice that this condition is not as strict as the threshold condition on the synthesis rates. 

Besides these main two conditions, high binding rate, low unbinding rate and sufficient stability of the complex are also needed as shown in Figure \ref{fig:singlescan}(e-g). This is consistent with the previous knowledge that the binding rate strongly influences the effectiveness of the toxin-antitoxin mechanism \cite{Hjalt92,Nordgren01}. Finally in Figure \ref{fig:singlescan}(i), when the protein degradation rate $\beta_{p}$ is very low (equivalent to having a half life of longer than $5$ min), $R$ decreases drastically and eventually drops to a constant around $4$, where no protein is being degraded. Thus, the proteolysis of the toxic protein contributes to the efficiency of the toxin-antitoxin mechanism for plasmid maintenance. The translation (protein synthesis) rate, on the other hand, has no effect on the ratio $R$ as expected (Figure \ref{fig:singlescan}(h)).

A few surprises arise in the parameter dependence studys. First of all, it is known that the antitoxin sRNA in plasmid number maintenance is very short-lived \cite{Gerdes88} (which is indeed true for many types of sRNA \cite{Brenner91,Stougaard81}), while the toxin mRNA has an unusually long half-life \cite{Gerdes88}. However, in the numerical study above we find when the degradation rate of the sRNA exceeds that of the template mRNA, in some range, the mechanism becomes actually less effective: $R$ decreases in Figure \ref{fig:singlescan}(d) after $\beta_{s}$ becomes larger than around $0.5$. However, we did not see this effect clearly in the random sampling of the parameter space in the following section. It is therefore possible this effect only applies to situations where some other parameters, for example the synthesis rates of the RNA molecules, take on certain values. 

Another surprise comes from the complex stability. In some well-studied cases, the binding of mRNA with sRNA leads to the rapid degradation of the complex \cite{Lenz06}. However, this cannot be the case here. As shown in \ref{fig:singlescan}(g), when the RNA complex is degraded at a rate higher than $0.8$ there is no transient toxin peak at all ($R=0$). This can be explained by the fact that mRNA must be released from the complex upon plasmid loss, which requires a sufficiently high concentration of the complex. If the complex degradation rate $\beta_{c}$ is too high, too few mRNA templates will be left for the translation of the toxin. Nevertheless, complex stability is an important parameter beyond this obvious feature, as will be shown by the second parameter dependence study conducted in the following section \ref{sec:rand} Figure \ref{fig:randscan}(b). 

\subsection{Random Sampling of the Parameter Space} \label{sec:rand}

We have seen that the dependence of the toxin-antitoxin circuit on the individual parameters exhibit expected but also surprising behaviors. However, due to the fact that in our single parameter scans only one parameter was varied at a time, while the other 9 are fixed, the generality of our conclusions is up for debate. To find the region of the 10-dimensional parameter space where the toxin-antitoxin mechanism works effectively, and to explore the joint conditions on the parameters, we randomly sample the parameter space.

We run a total of 4025 simulations of the toxin-antitoxin dynamics with randomly generated parameters. The only restrictions imposed on the random parameter sampling are $\alpha_{m} > \beta_{m}$ and $\alpha_{s} > \beta_{s}$, to make sure that average molecule numbers exceed $1$ and the differential equations approach is appropriate. The algorithm generates random values for each of the 9 parameters ($g=6$ remains fixed) within the range given in Table \ref{tab:Tablepar} and with the aforementioned restriction, at an appropriate sampling resolution. The resolution for $\alpha_{m}$, $\beta_{m}$, $\alpha_{s}$, $\beta_{s}$ are chosen such that the log of the ratios $\alpha_{m}/\alpha_{s}$ and $\beta_{m}/\beta_{s}$ are uniformly distributed. The sampling is therefore random and to a large degree uniform, or log-uniform. 

\begin{figure}[!thbp]
\centering
\captionsetup[subfigure]{justification=centering}
\includegraphics[width=1\textwidth]{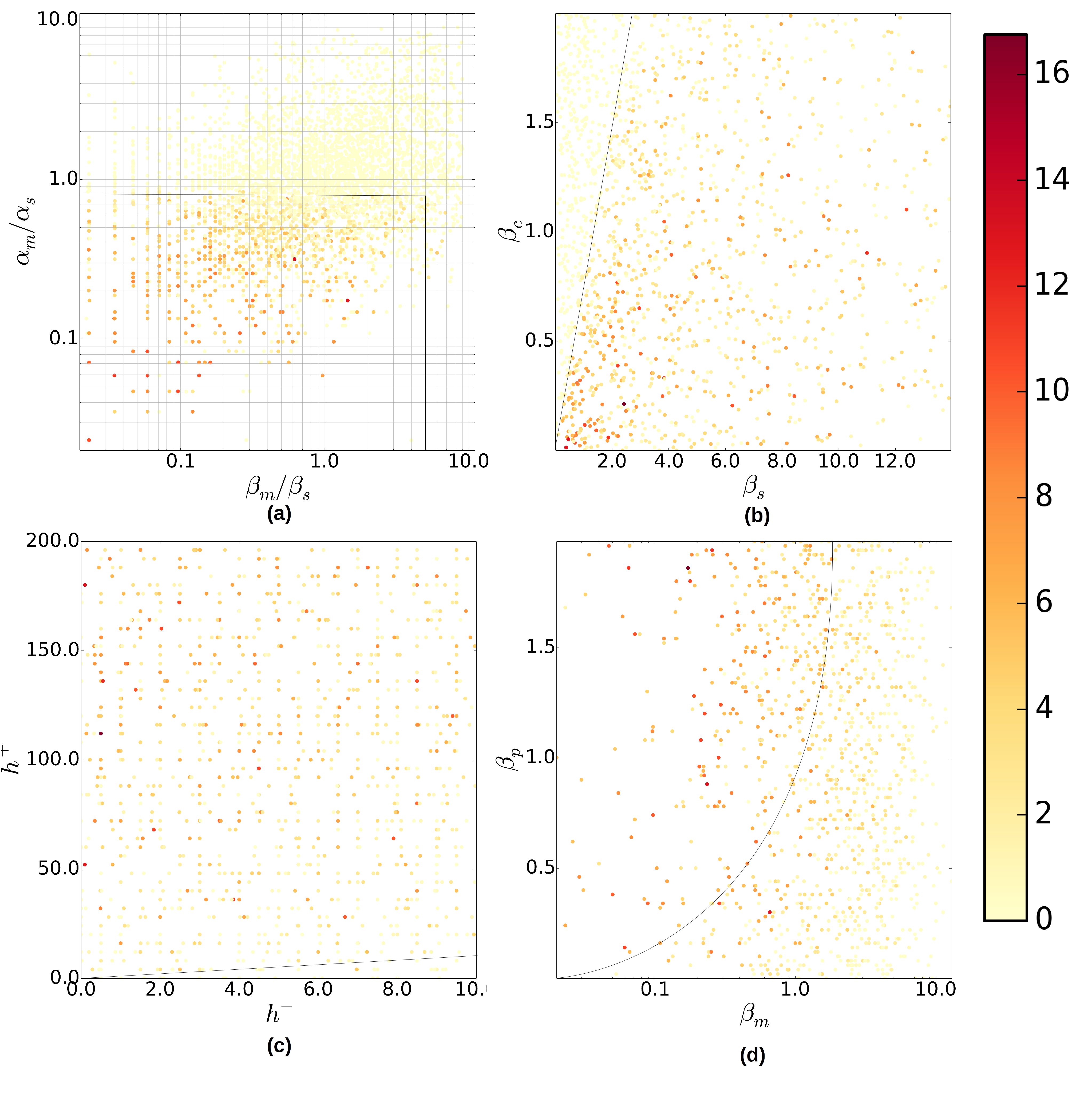} 
\caption{Parameter dependence found by random parameter sampling: The color gradient shows the magnitude of the protein peak fold-increase $R$. In Figure (a), the results are projected onto $\alpha_{m}/\alpha_{s}$ vs. $\beta_{m}/\beta_{s}$ on a log-log scale. In (b) the simulation results that fall within the subregion $\alpha_{m}/\alpha_{s} < 0.8$ and $\beta_{m}/\beta_{s} < 4$ (marked by lines in Figure (a)) are projected onto $\beta_{c}$ vs. $\beta_{s}$. In (c,d) the parameter sets are further restricted to the subregion $\alpha_{m}/\alpha_{s} > 0.8$ , $\beta_{m}/\beta_{s} > 4$ and $\beta_{c}/\beta_{s} < 2/3$ (marked by a dark line in Figure (b)) and the results are projected onto $h^{+}$ vs. $h^{-}$ in Figure (c) , and $\beta_{p}$ vs. $\beta_{m}$ in Figure (d), respectively. The color scale values are taken to be $log_{1.4}R$ for ease of viewing. Dark lines mark the rough division between the region with more high peaks and the region with fewer. In (c) and (d) the dark lines mark $h^{+} = h^{-}$ and $\beta_{p} = \beta_{m}$ respectively. \label{fig:randscan}}
\end{figure}

The value of the protein concentration peak fold-increase $R$ for each simulation is recorded and projected onto the space spanned by $\alpha_{m}/\alpha_{s}$ and $\beta_{m}/\beta_{s}$ (Figure \ref{fig:randscan}(a)). The observations obtained by random sampling are consistent with the results of the single parameter dependence study. The results confirm that $\alpha_{m} < \alpha_{s}$ is indeed a strong criterion for the existence of a high protein concentration peak. Figure \ref{fig:randscan}(a) shows when $\alpha_{m}/\alpha_{s} > 0.8$ there is almost no protein concentration peak after the loss of plasmids. $\beta_{m} < \beta_{s}$ is also a very important criteria, but exceptions are allowed: The protein level decreases on average when $\beta_{m} > \beta_{s}$, nevertheless, there are distinct protein concentration peaks beyond $\beta_{m}/\beta_{s} > 1$. However, the majority of the simulations with $\beta_{m}/\beta_{s} > 1$ exhibit no peaks. These exceptions could be due to some other competing effects from the dynamics of the mRNA-sRNA complex or from the dynamics of the RNA molecules. 

\begin{table}[hb]
\footnotesize
\begin{tabular}{|p{3.5cm}|p{1.2cm}|p{1.2cm}|p{1.2cm}|p{1.2cm}|p{1.2cm}|p{1.2cm}|p{1.2cm}|p{1.2cm}|p{1.2cm}|}\hline
Parameter name & $\alpha_{m}$ & $\beta_{m}$ & $\alpha_{s}$ & $\beta_{s}$ & $h^{+}$ & $h^{-}$ & $\beta_{c}$ & $\alpha_{p}$ & $\beta_{p}$\\
\hline
Maximum Value & 20 & 14 & 20 & 14 & 200 & 10 & 2 & 30 & 2\\
\hline 
Miminum Value & 0.001 & 0.001 & 0.001 & 0.001 & 0.1 & 0.001 & 0.001 & 0.01 & 0.001\\
\hline 
Sampling Resolution & 0.0005 & 0.0005 & 0.0005 & 0.0005 & 4 & 0.005 & 0.001 & 0.9 & 0.02\\
\hline 
\end{tabular}
\\
\caption{Parameter space being sampled and the rates of sampling. 
\label{tab:Tablepar}}
\end{table}

Next, we focus on the cases that satisfy $\alpha_{m}/\alpha_{s} < 0.8$ and $\beta_{m}/\beta_{s} < 4$, and plot those as a function of the degradation rate of the complex $\beta_{c}$ and that of the sRNA $\beta_{s}$. This will eliminate toxin-antitoxin dynamics with parameter combinations that are inefficient in producing a transient peak due to the criteria discussed above, allowing new effects to be discovered more easily. Figure \ref{fig:randscan}(b) shows that no peaks occur for $\beta_{c} \gtrsim \beta_{s}$, in contrast to the rest of the region. In fact, the peaks are visually prominent only when $1.5*\beta_{c} \lesssim \beta_{s}$.
A likely explanation is as follows: when $\beta_{c}$ is low, there is a higher level of total mRNA molecules in bound form both before and after the plasmid loss. Because the DNA gene copies are removed when the plasmids are lost, the transient increase of protein synthesis is almost entirely due to the mRNA released from the complex. Therefore, the complex degradation (as opposed to mRNA degradation) is the dominant reason for total mRNA loss. At the same time, lower $\beta_{c}$ does not increase mRNA expression before plasmid loss, but merely increases the amount of bound mRNA molecules, and will therefore not lead the plasmid-containing cell to ``poison itself''. If this loss of mRNA due to complex degradation is slower than the loss of sRNA (the loss of sRNA is equal to the gain of the free mRNA when the plasmids are lost), the toxic protein can be synthesized from the surplus free mRNA. Therefore, if antitoxin sRNA is less stable than the total mRNA, there will be a transient peak of mRNA, and correspondingly one for the toxic protein, after plasmid loss. However, when a higher level of free mRNA is present before plasmid loss, due to faster degradation of sRNA, i.e. when $\beta_{s}$ is large, $R$ is in turn negatively affected. Considering this negative impact, $\beta_{c} \lesssim \beta_{s}$ therefore does not always guarantee high peaks, which is shown from a wide range of variations in $R$ within this region, and there is no increase in $R$ when $\beta_{c}$ is increasingly small compared to $\beta_{s}$. 

It is surprising that the stability of the complex plays such an important role in the toxin-antitoxin mechanism. It is secondary only to the threshold condition on the synthesis rates and the differential stability of the RNA molecules. This highlights the crucial role played by the RNA complex in the dynamics. In connection to the single parameter study conducted before, this demonstrates that the two known effects of unstable sRNA \cite{Gerdes88} and unstable RNA complexes \cite{Lenz06} need to be understood in relation to each other and not in isolation. Unstable RNA complexes are only useful for the functioning of the toxin-antitoxin system if the sRNA is more unstable than the RNA complex. 

Next, we restrict our set of parameter combinations further, by imposing $\beta_{c}/\beta_{s} \lesssim 2/3$, $\alpha_{m}/\alpha_{s} < 0.8$ and $\beta_{m}/\beta_{s} < 4$. Figure \ref{fig:randscan}(c) shows that for a prominent protein concentration peak to occur, $h^{-}/h^{+}$ should not be larger than one. Projections onto $\beta_{p}$ and $\beta_{m}$ show that when $\beta_{p}$ is close to $0$ there is no major peak (Figure \ref{fig:randscan}(d)). This means that by making extremely stable toxic proteins we cannot increase the relative increase in protein concentration. It also shows generally $\beta_{m} < \beta_{p}$ result in less pronounced peaks than $\beta_{m} > \beta_{p}$. This is usually satisfied in real situations, since mRNA molecules are typically less stable than proteins. 

The fact that high peaks do not occur in the subregion with $h^{-}/h^{+} > 1$ tells us that the binding rate needs to be higher than the unbinding rate, i.e. the binding needs to be strong. However, it also cannot be infinitely strong, i.e. irreversible, because in that case there will be no free mRNA available for protein synthesis after plasmid loss.

Consistent with the results from the single parameter dependence study, the synthesis rate $\alpha_{p}$ does not have an obvious effect on the protein concentration peak. This shows that the protein concentration peak is mostly a result of the dynamics between the mRNA, sRNA and the RNA complex, characterized for example by relations between $\beta_{m}$ and $\beta_{s}$, $\alpha_{m}$ and $\alpha_{s}$, $\beta_{c}$ and $\beta_{s}$, and not of a high synthesis rate of the protein. This also emphasizes the importance of understanding sRNA regulation purely from the point of view of RNA dynamics instead of protein dynamics, which is traditionally viewed as playing the dominant role in cellular regulation.

To summarize, by randomly sampling the parameter space as well as performing a single parameter dependence study, we determine that the conditions for an efficient toxin-antitoxin mechanism are as follows:
\begin{enumerate}
\item The synthesis rate of mRNA should be lower than the synthesis rate of sRNA;
\item The degradation rate of mRNA should be relatively low, compared to the degradation rate of sRNA;
\item The stability of the complex should be higher than the stability of the sRNA antitoxin;
\item A high affinity and irreversibility (but not complete irreversibility) in RNA complex formation and some degrees of protein instability are also helping factors, but are of less significance.
\item Protein synthesis rate does not contribute to the forming of a transient protein concentration peak. 
\end{enumerate}
Despite the nonlinear nature of our system, by individually controlling the available parameters, a genetic circuit could be engineered to produce specific effects, such as a higher increase in toxin concentration after the loss of plasmids for an effective duration (from a few minutes up to an hour). 

\subsection{Analytical approximation to the transient protein concentration peak}

Using a simplified version of the differential equations after plasmid loss, an analytical expression for the transient peak can be obtained, which qualitatively describes the dynamics. For this approximation, we assume that after the plasmid copies are lost, all mRNA molecules are immediately available in free form. That is, the number of free mRNA molecules after plasmid loss equals the sum of the mRNA steady state concentration $m^{*}$ and the complex steady state concentration $c^{*}$. The differential equations which describe the dynamics are as follows:
\begin{equation} \label{eq:simp}
\begin{aligned}
 \dot m & = \beta_{m} \cdot m \\
 \dot p & = \alpha_{p} \cdot m - \beta_{p} \cdot p.
 \end{aligned}
\end{equation}
The initial conditions and integration constants required to solve the differential equations (\ref{eq:simp}) are provided by the steady state concentrations before and after the loss of plasmids, respectively. Explicit expressions for mRNA and protein concentration as a function of time can then be obtained:
\begin{subequations}
 \begin{align}
 m & = m_{2}^{*} + (m^{*} + c^{*} - m_{2}^{*})e^{-\beta_{m}t} \\
 p & = p_{2}^{*} - \frac{(m^{*} + c^{*} - m_{2}^{*})\alpha_{p}}{\beta_{m} - \beta_{p}}e^{-\beta_{m}t} + (p^{*} - p_{2}^{*} + \frac{(m^{*} + c^{*} - m_{2}^{*})\alpha_{p}}{\beta_{m} - \beta_{p}})e^{-\beta_{p}t},
 \end{align}
\end{subequations}
where $m^{*}$, $c^{*}$ and $p^{*}$ are the steady state concentrations for mRNA, complex and protein before plasmid loss, and $m_{2}^{*}$ and $p_{2}^{*}$ are the steady state concentrations after plasmid loss (see Section \ref{sec: secss}). In the case that all plasmid copies are lost, the steady state concentrations $m_{2}^{*}$ and $p_{2}^{*}$ are both zero and the above expressions can be simplified to:
\begin{subequations}
 \begin{align}
 m & = m^{*} e^{-\beta_{m}t} \\
 p & = - \frac{(m^{*} + c^{*})\alpha_{p}}{\beta_{m} - \beta_{p}}e^{-\beta_{m}t} + (p^{*} + \frac{(m^{*} + c^{*})\alpha_{p}}{\beta_{m} - \beta_{p}})e^{-\beta_{p}t}. \label{eq:anl2b}
 \end{align}
\end{subequations}

\begin{figure}[h]
\centering
\captionsetup[subfigure]{justification=centering}
\includegraphics[width=0.7\textwidth]{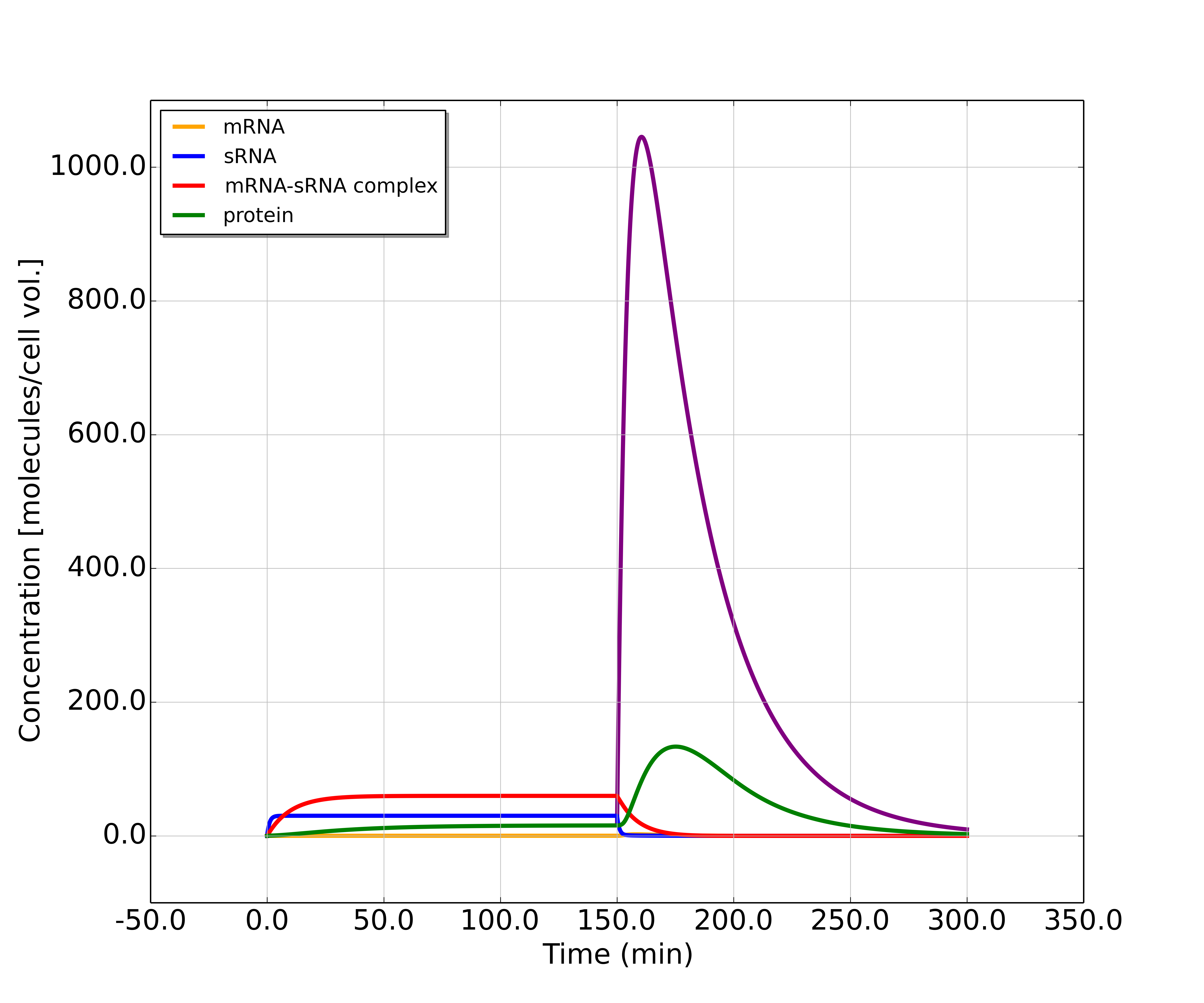}
\caption{Analytical approximation of the dynamics after plasmid loss: With the same parameter values as in Figure \ref{fig10real}, Solution \ref{eq:anl2b} gives a very large peak in approximated protein concentration (purple) compared to the actual simulated protein concentration change (green). It should be noted that this approximation considers no reaction time delay after plasmid loss. \label{fig:anl2}}
\end{figure}

These expressions qualitatively describe the transient peak of the protein concentration after plasmid loss, however the height of the peak is strongly overestimated (Fig \ref{fig:anl2}). In the full model, mRNA is released slowly from the complex, which reduces the peak height. This behavior can be mimicked within the approximation by the sudden release of only a fraction of the mRNAs (replacing $m^*$ by an effective mRNA concentration somewhere between $m^*$ and $c^*$). 

Despite these quantitative shortcomings of the approximation, it is useful to understand some properties of the dynamics. For example, one can easily see from the solution (Eq.\ref{eq:anl2b}), that the translation rate does not affect the peak fold-increase $R$, because both terms are proportional to $\alpha_p$, while the steady state concentration before plasmid loss, $p^{*}$, is also proportional to $\alpha_{p}$. Thus, $R$ being the ratio of the two is independent of $\alpha_{p}$. This result should also be true for the full dynamics since the analytical approximation used here is only considering the RNA dynamics and assumptions about the protein dynamics have not been made. The dependence on $\beta_{m}-\beta_{p}$ is also important: An increase of this quantity is equivalent to an increase of $p^{*}$, consistent with the tendency observed in the random parameter sampling %in section \ref{sec:rand} 
(Figure \ref{fig:randscan}d).

\subsection{Introduction of a Competitor mRNA to Increase Toxin Levels}

Regulation by small RNAs typically displays crosstalk with multiple mRNAs under the control of the same sRNA \cite{Levine08}. In the case of toxin-antitoxin systems, this can be exploited as an antibacterial strategy \cite{Williams,Faridani}. By inducing a gene encoding a competitor mRNA (on the same or another plasmid or on the chromosome) that can bind to the antitoxin sNRA and hence allows toxin mRNA to exist in free form, a similar increase in toxin concentration can be induced as with plasmid loss, which may also lead to the killing of the cell.
To describe the dynamics of this scenario, the equations from above are extended to include a second type of mRNA, the competitor, and the corresponding mRNA-sRNA complex. The full dynamics is then described by the following equations:
\begin{equation} \label{eq:extra}
\begin{aligned}
 \dot m & = \alpha_{m} \cdot g - \beta_{m} \cdot m - h^{+} \cdot m \cdot s + h^{-} \cdot c \\
 \dot s & = \alpha_{s} \cdot g - \beta_{s} \cdot s - h^{+} \cdot m \cdot s + h^{-} \cdot c - k^{+} \cdot m_{2} \cdot s + k^{-} \cdot c_{2} \\
 \dot c & = h^{+} \cdot m \cdot s - h^{-} \cdot c - \beta_{c} \cdot c \\
 \dot p & = \alpha_{p} \cdot m - \beta_{p} \cdot p \\
 \dot m_{2} & = \alpha_{2} \cdot g - \beta_{2} \cdot m_{2} - k^{+} \cdot m_{2} \cdot s + k^{-} \cdot c_{2} \\
 \dot c_{2} & = k^{+} \cdot m_{2} \cdot s - k^{-} \cdot c_{2} - \beta_{c_{2}} \cdot c_{2},
\end{aligned}
\end{equation}
where $m_{2}$ and $c_{2}$ denote the concentrations of the competitor mRNA and the complex it forms with an sRNA molecule, respectively. $k^{+}$, $k^{-}$ and $\beta_{c_{2}}$ are the binding rate, unbinding rate and the degradation rate of this new complex. 

\begin{figure}[!thbp]
\centering
\captionsetup[subfigure]{justification=centering}
\includegraphics[width=0.9\textwidth]{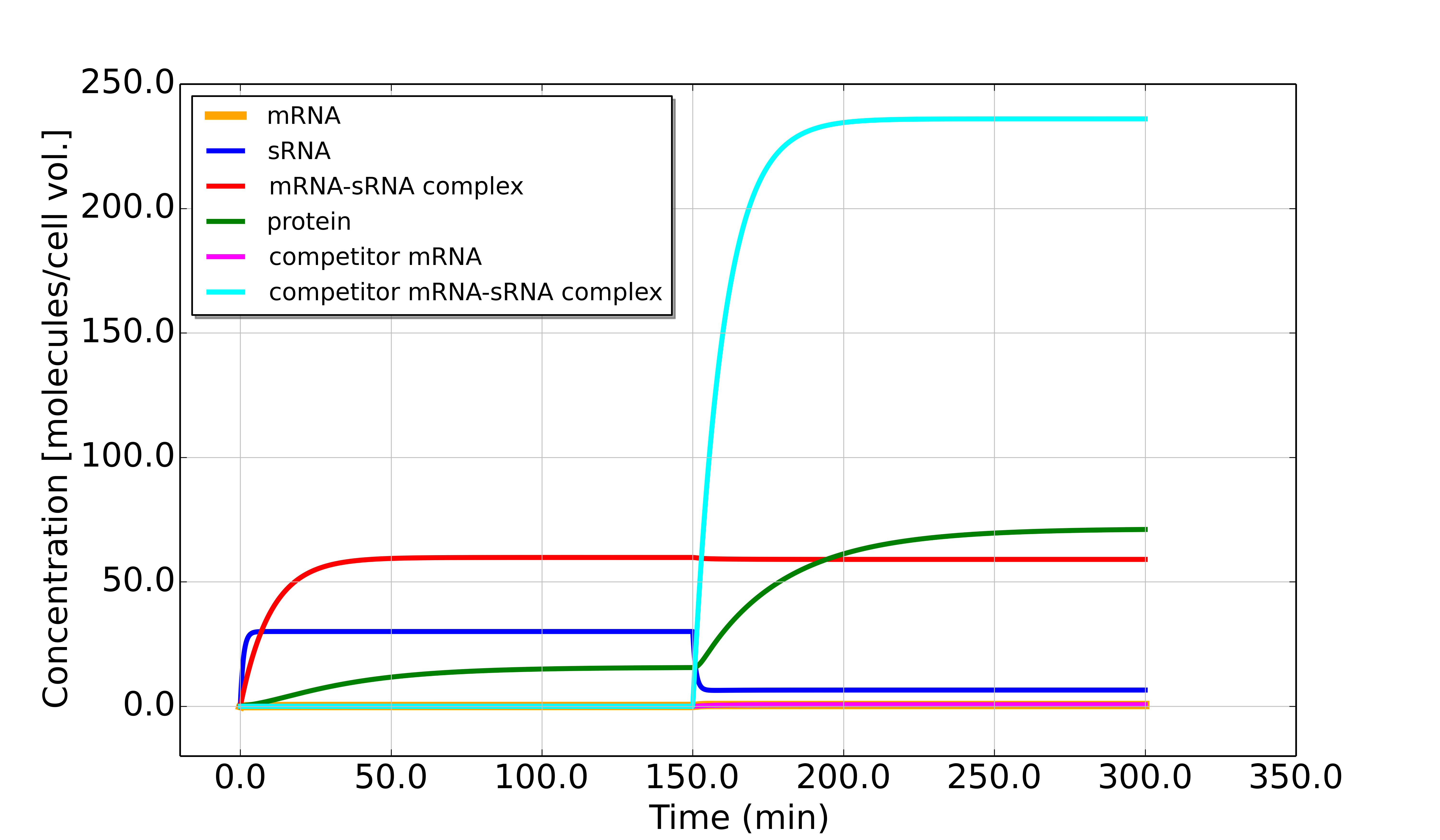}
\caption{Introduction of a competitor mRNA: Concentrations of sRNA, mRNA, mRNA-sRNA complex, competitor mRNA, complex formed by competitor mRNA and sRNA and protein as a function of time. Note the non-zero steady state concentrations after the triggering event resulting from the new dynamics. The parameters in equation (\ref{eq:extra}) are as follows: the first 10 parameters are the same as in Figure \ref{fig:singlescan} . Parameter values for competitor mRNA are chosen from within the experimental values of mRNA synthesis rate and life time, with synthesis rate $\alpha_{2} = 4.0$ and degradation rate $\beta_{2} = 0.6$. The competitor mRNA-sRNA complex binding and unbinding rates are $k_{+} = 60$ and $k_{-} = 1.0$. The competitor RNA complex degradation rate is $\beta_{c_{2}} = 0.1$. Toxin mRNA level is found to be low throughout the simulation. \label{fig:extra}}
\end{figure}

We assume that the competitor is induced at the time $t = 150$ min. Then, from $t = 0$ to $150$ min, the dynamics is the same as Equation (\ref{eq:1}). At $t = 150$ min, Equation (\ref{eq:extra}) are being integrated, which corresponds to the competitor gene being turned on. With zero initial concentrations for all components (at $t = 150$ min, new variables $r = v = 0$), an example simulation is shown in Figure \ref{fig:extra} .

A transient protein concentration peak can be observed with some specific combinations of parameters, but the majority of the simulations we run show no transient peak, as in Figure \ref{fig:extra} . However, a non-zero steady state toxin concentration exists after induction of the competitor which can be effective in killing the host cell, since the mRNA encoding the toxic protein is still produced from the plasmid. Thus, in contrast to the case considered before, here the toxin-antitoxin system does not perform its key function in a transient dynamical fashion, but rather in the steady state, with the competitor either induced or not induced. Therefore, we define a new fold-increase parameter $\widetilde{R}$ of the toxin concentration as the ratio between the steady states  of the protein concentration before and after the synthesis of the competitor RNA.

Similar to the procedure in Section \ref{sec: secss}, the new steady state concentration for the mRNA $\widetilde{m_{2}^{*}}$ turns out to be a positive root to the cubic equation (\ref{eq:cubic}):
\begin{equation} \label{eq:cubic}
\begin{aligned}
\alpha_{2}g(1+\frac{k^{-}}{\beta_{c_{2}}})m^{2} - \frac{\beta_{s}}{\beta_{2}h^{+}}\left\{ \left[\beta_{2}(1+\frac{k^{-}}{\beta_{c_{2}}})- \frac{k^{+}}{h^{+}} \beta_{m} (1+\frac{h^{-}}{\beta_{c}})\right]m + \frac{k^{+}}{h^{+}}\alpha_{m}g(1+\frac{h^{-}}{\beta_{c}})\right\} & \\
\times \left\{ - \frac{\beta_{m}}{\beta_{s}}h^{+} m^2 - \beta_{m}(1+\frac{h^{-}}{\beta_{c}})m - \frac{(\alpha_{s}- \alpha_{m} - \alpha_{2})gh^{+}}{\beta_{s}}m + \alpha_{m}g(1+\frac{h^{-}}{\beta_{c}}) \right\} = 0
\end{aligned}
\end{equation}
After algebraic manipulations, it can be written in canonical form as follows. 
\begin{equation} 
A m^{3} + B m^{2} + Cm + D = 0
\end{equation}
where $A$, $B$, $C$ and $D$ are functions of the rates.

An analytical formula for the solutions of a cubic equation can always be given explicitly, but the expressions are excessively lengthy and hence not shown here. With a root finding algorithm such as the bisection method one can easily find the three roots to the cubic equation numerically. Using values similar to the experimental values, we found that even in cases where there is more than one positive root, we can still pick out the correct solution because usually one of the two positive roots corresponds to an unrealistically large concentration.

The analytical solution to Equation (\ref{eq:cubic}) is then compared to the numerically generated steady state concentrations after integration, and very good agreement (up to floating point error) is found. Once we obtain the steady state mRNA concentration $\widetilde{m_{2}^{*}}$, then following $\widetilde{p_{2}^{*}} = \widetilde{m_{2}^{*}} \cdot \alpha_{p}/\beta_{p}$ one can solve for the steady state protein concentration. The ratio $\widetilde{p_{2}^{*}}$ / $p^{*}$, where $p^{*}$ is given by Equation (\ref{eq:pstar}), gives an analytical value for the fold-increase $\widetilde{R}$. This is how we could theoretically obtain an analytical formula for $\widetilde{R}$ in the case of competitive RNA binding, which is just as effective in killing the host cell at some triggering event as the original toxin-antitoxin mechanism. 

\begin{figure}[!thbp]
\captionsetup[subfigure]{justification=centering}
\centering
\includegraphics[width=1\textwidth]{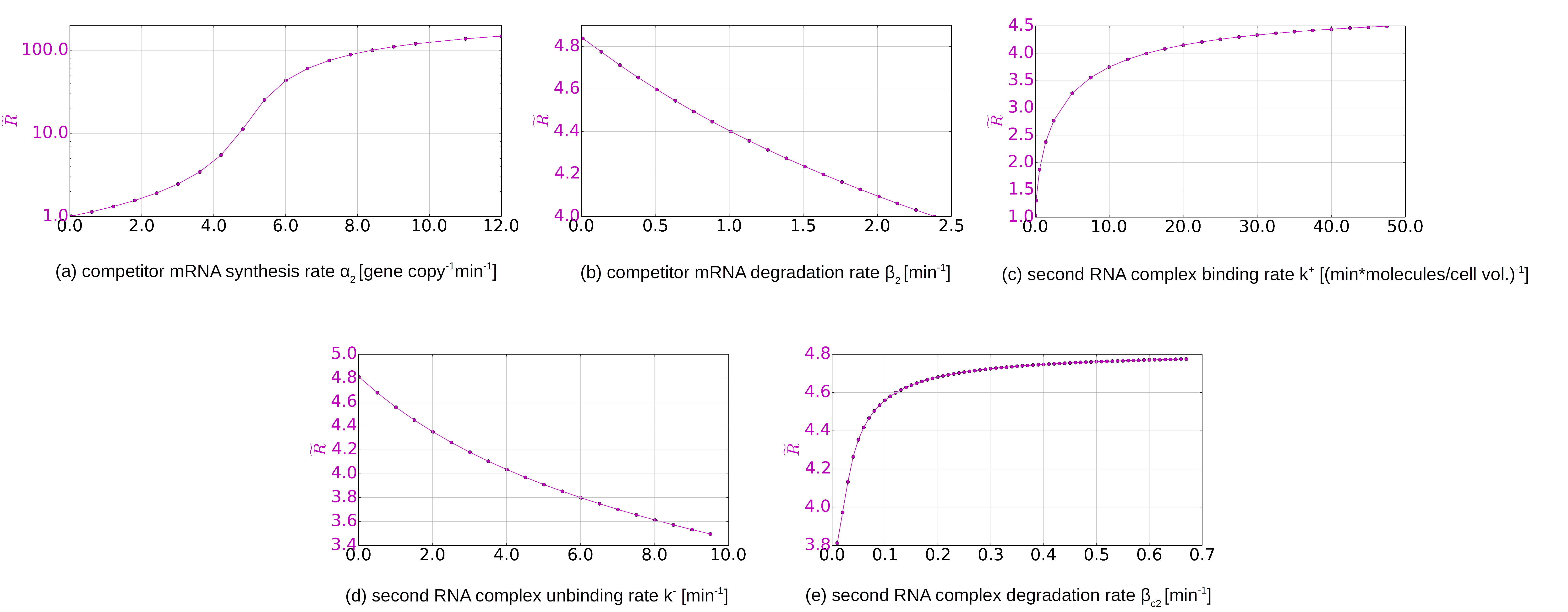}
\caption{\label{fig:compscan} Dependence on the parameters of the competitor: each new parameter of the system is increased along the $x$ axis. Toxic protein concentration peak fold-increase $\widetilde{R}$ is plotted on the y axis for each run corresponding to the value the new parameter takes on. The default or fixed values of the parameters are the same as in Figure \ref{fig:extra}.}
\end{figure}

Figure \ref{fig:compscan} shows the single parameter variation response plots generated with the same method as used in Figure \ref{fig:singlescan} .
We observe that as soon as the synthesis of the competitor is turned on, i.e. $\alpha_{2} > 0 $, there is a substantial fold-increase in the toxin concentration, i.e. $\widetilde{R} > 1$. As $\alpha_{2}$ becomes large, the increase in $\widetilde{R}$ is diminished because all sRNA molecules are bound. Changing the degradation rate of the competitor mRNA $\beta_{2}$ results in small variation in the toxin fold-increase response $\widetilde{R}$, meaning that for the default parameter combination, there are very few free competitor mRNA molecules and most are bound in complexes. The degradation rate of the competitor mRNA-sRNA complex $\beta_{c_{2}}$ positively affects $\widetilde{R}$, because it increases the destruction of the antitoxin in bound state, allowing more target mRNA to be in free form available for protein synthesis. 

The parameter variation in Figure \ref{fig:compscan} shows that the synthesis rate of the new competitor mRNA typically plays the dominant role compared to other new parameters, a general feature of cross-talk in sRNA regulation \cite{Levine08}.

\section{Conclusion and Summary}

We have studied the toxin-antitoxin mechanism for plasmid maintenance through post-segregational killing using a theoretical model. The model extends previously studied models for sRNA-dependent regulation that have considered the limiting cases of rapid equilibrium or irreversible binding, but have thereby largely ignored the nonlinear nature of the RNA complex binding and the dynamics of the complex itself. Here we have taken the full dynamics into account and given an analytical solution for the steady state concentrations (where the plasmid copy number remains constant). We simulated the dynamics before and after plasmid loss by numerical integration and showed that the toxin-antitoxin system performs its function by inducing a transient peak in the toxin concentration, which must exceed a threshold for host killing. The formation of this peak depends on the release of the mRNA template of the protein toxin from the mRNA-sRNA complex. The accumulation of the mRNA molecules is only possible when sRNA is degraded at a faster rate than the mRNA. 

Using two kinds of parameter variations to study the parameter dependence of the system (individual parameter are varied in isolation, and all parameters are varied simultaneously in a random manner within a given range), we can draw a number of conclusions with regards to which ones of the $10$ system parameters have significant effects on the system, and why when they take on certain ranges of values, the system functions more optimally than in other cases. In general we have found that, for an efficient toxin-antitoxin mechanism, the synthesis rate of toxin's mRNA template should be lower than that of the sRNA antitoxin, the mRNA template should be more stable compared to the sRNA antitoxin, and the mRNA-sRNA complex should be more stable than that of the sRNA antitoxin. Analytically approximating the protein peak by allowing all mRNA to be released at once gives us an analytical expression for the peak, which despite overpredicting its height nonetheless gives qualitative insights that are consistent with previous numerical observations. 

Finally, we also studied the possibility of inducing the toxic protein with a competitor mRNA, which sequesters the sRNA antitoxin. Such a mechanism has been proposed as an antibacterial strategy \cite{Williams,Faridani}. Here the effectiveness in killing the host cell depends on the ratio between the steady states of the toxin concentration before and after introducing the competitor RNA. An analytical solution can be given for this ratio by solving a cubic equation. We performed a single-parameter variation study for this scenario and found that the dominant parameter dependence here is on the synthesis rates. A sufficiently high synthesis rate results in all antitoxin being sequestered and the synthesis of protein toxin. In such a way the cross-talk inherent in sRNA regulation could be utilized towards inducible killing of the cells. 

\clearpage

\section{Bibliography}
\bibliographystyle{unsrt}
\bibliography{biblio}
\end{document}